\documentclass{article}

\usepackage[nonatbib,preprint,dblblindworkshop]{neurips_2025} 
\usepackage[numbers]{natbib}
\usepackage[utf8]{inputenc} 
\usepackage[T1]{fontenc}    
\usepackage{hyperref}       
\usepackage{url}            
\usepackage{booktabs}       
\usepackage{amsfonts}       
\usepackage{amsmath}
\usepackage{nicefrac}       
\usepackage{microtype}      
\usepackage{xcolor}         
\usepackage{algorithm}
\usepackage{algpseudocode}
\usepackage{graphicx}

\title{GO-Diff: Data-free and amortized global structure optimization}

\author{
  Nikolaj Rønne\textsuperscript{\textdagger}, Tejs Vegge and Arghya Bhowmik \\
  Technical University of Denmark \\
  CAPeX Pioneer Center for Accelerating P2X Materials Discovery \\
  {}\textsuperscript{\textdagger}\texttt{niron@dtu.dk}
}

\begin{document}

\maketitle

\begin{abstract}
We introduce GO-Diff, a diffusion-based method for global structure
optimization that learns to directly sample low-energy atomic
configurations without requiring prior data or explicit
relaxation. GO-Diff is trained from scratch using a Boltzmann-weighted
score-matching loss, leveraging only the known energy function to
guide generation toward thermodynamically favorable regions. The
method operates in a two-stage loop of self-sampling and model
refinement, progressively improving its ability to target low-energy
structures. Compared to traditional optimization pipelines, GO-Diff
achieves competitive results with significantly fewer energy
evaluations. Moreover, by reusing pretrained models across related
systems, GO-Diff supports amortized optimization — enabling faster
convergence on new tasks without retraining from scratch. 
\end{abstract}

\section{Introduction}
The potential energy surface (PES) is a high-dimensional, non-convex
function that encodes the stability of atomic configurations by
mapping atomic positions to their associated potential
energy. Exploring this surface to identify low-energy (and thus
stable) structures is a fundamental challenge in computational
materials science, chemistry, and catalysis.\cite{oganov2019} This task, often referred
to as global structure optimization, underpins applications ranging
from the discovery of new catalytic surfaces to the design of
functional materials. 

Traditional global optimization methods — such as random structure
search (RSS)\cite{pickard2011}, basin-hopping\cite{wales1997}, genetic algorithms\cite{deaven1995}, and simulated
annealing\cite{kirkpatrick1983} — rely on local relaxation with gradient-based optimizers to
identify nearby minima. While effective, these methods are
computationally expensive due to the many energy and force evaluations
required, and their reliance on local optimization can limit the exploration of complex energy
landscapes. Machine learning interatomic potentials\cite{bartok2010,schutt2018,schutt2021,batzner2022,batatia2022}, whether
pre-trained\cite{deng2023,batatia2023} or learned
on-the-fly\cite{bisbo2020,ronne2022}, can reduce the cost of local optimization. However, they
require carefully selected training data to capture relevant minima;
otherwise, the search risks getting trapped in self-reinforcing local
minima (by only gathering new data near known regions of the PES) far from the true global optimum. 

Score-based diffusion models\cite{song2021} have shown promise for structure
generation in molecular\cite{hoogeboom2022,xu2023,cornet2024} and materials science\cite{xie2022,jiao2023,ronne2024,zeni2025}. However, applying them
to global optimization tasks is challenging: the goal is to sample
rare, low-energy configurations corresponding to the global minimum of
a PES, but the data distribution over such
structures is typically unknown or inaccessible. This makes it
difficult to train models that prioritize physically meaningful,
low-energy samples. 

In this work, we introduce GO-Diff, a diffusion-based framework for
global structure optimization that operates without prior data by
leveraging the known energy function during training. GO-Diff learns
to generate low-energy configurations without explicit relaxation
using a Boltzmann-weighted score-matching loss combined with iterative
self-sampling and annealed training. 

Our key contributions are:
\begin{enumerate}
\item \textbf{A data-free generative optimization method} that
  directly samples minima of the potential energy surface. 
\item \textbf{A Boltzmann-weighted loss with annealing} to guide
  sampling toward low-energy regions while maintaining exploration. 
\item \textbf{Amortized optimization} through transfer of pretrained
  models across related systems. 
\item \textbf{Empirical evidence} of superior sample efficiency
  compared to classical search methods.
\end{enumerate}

GO-Diff departs from prior Boltzmann samplers (see SI) by using a direct
Boltzmann-weighted score-matching loss that avoids force evaluations
and Monte Carlo estimates, and by demonstrating amortized optimization
through transfer of pretrained models across related systems. 

\section{Methods}

\paragraph{Training loop.}
GO-Diff optimizes atomic structures by training a diffusion model to
generate low-energy configurations without requiring data or local
relaxation. It operates in a self-sampling loop: the model generates
atomic structures via reverse-time diffusion, evaluates their
energies, and uses the resulting samples to refine itself. 

A replay buffer $\mathcal{B} = \{(\mathbf{x}^{(i)}_0, E^{(i)})\}$ stores generated
configurations $\mathbf{x}^{(i)}_0$ along with their energies $E^{(i)}$. The
buffer is initially seeded with $N$ samples from an untrained
model. At each iteration, the current model samples $N$ new structures
from the reverse stochastic differential equation (SDE), which are then evaluated using an 
energy function. These are merged with existing buffer entries.

This iterative process enables GO-Diff to learn directly from its own
generations, progressively focusing on thermodynamically favorable
regions of the energy landscape. 

\paragraph{Boltzmann-weighted score matching.}
We wish to sample from the Boltzmann distribution\cite{noe2019} at
temperature $T$
\begin{equation}
\pi_T(\mathbf{x}) = \frac{\exp(-E(\mathbf{x})/T)}{Z_T},
\qquad
Z_T = \int \exp(-E(\mathbf{x})/T)\,d\mathbf{x},
\end{equation}
where $E(\mathbf{x})$ is the potential energy of configuration $\mathbf{x}$.

If we had direct i.i.d.\ samples $\mathbf{x}_0 \sim \pi_T$, the denoising score
matching objective for training a score network $s_\theta(\mathbf{x}_t,t)$
would follow standard denoising score matching\cite{song2020}
\begin{equation}\label{eq:dsm}
\mathcal{L}_\theta = \mathbb{E}_{t \sim \mathcal{U}(0, 1)} \left[
  \lambda(t) \, \mathbb{E}_{\mathbf{x}_0 \sim \pi_T, \,
    \mathbf{x}_t \sim p_{t|0}(\mathbf{x}_t | \mathbf{x}_0)} \left\|
  s_\theta(\mathbf{x}_t, t) - \nabla_{\mathbf{x}_t} \log
  p_{t|0}(\mathbf{x}_t | \mathbf{x}_0) \right\|_2^2 \right], 
\end{equation}
where $s_\theta(\mathbf{x}_t, t)$ is the score model, $\mathbf{x}_t$ is
a noisy version of $\mathbf{x}_0$ and $p_{t|0}$ denotes the marginal of the forward SDE
conditioned on the initial data point usually referred to as the
transition kernel. 

Because direct samples from $\pi_T$ are unavailable, we approximate
expectations with samples $\mathbf{x}_0\sim q$, where $q$ denotes the empirical
buffer distribution. From importance
sampling (IS)\cite{hesterberg1995,agapiou2017a}, we have that for any
integrable $g$, 
\begin{equation}
\mathbb{E}_{\pi_T}[g(\mathbf{x})] = \mathbb{E}_{q} \left[ g(\mathbf{x})\frac{\pi_T(\mathbf{x})}{q(\mathbf{x})} \right].
\end{equation}

In practice, $q$ is not known analytically and we only store a buffer
$\mathcal{B}$ of evaluated configurations. 
From self-normalized importance sampling
(SNIS)\cite{agapiou2017a}, we have that the expectation can be approximated by
\begin{equation}
\mathbb{E}_{\pi_T}[g(\mathbf{x})]
\;\approx\; \sum_{i} w(E^{(i)}) \, g(\mathbf{x}^{(i)}),
\end{equation}
where the weights are the Boltzmann weights normalized over the buffer
samples
\begin{equation}
w(E) = \frac{\exp(-E/T)}{\sum_{E^{(i)} \in \mathcal{B}} \exp(-E^{(i)}/T)}.
\end{equation}


Substituting $g = \left\|  s_\theta(\mathbf{x}_t, t) - \nabla_{\mathbf{x}_t} \log p_{t|0}(\mathbf{x}_t | \mathbf{x}_0) \right\|_2^2$ from Eq.~\ref{eq:dsm} yields the
\emph{Boltzmann-weighted score matching} loss
\begin{equation}
\mathcal{L}^{\text{Boltzmann}}_\theta = \mathbb{E}_{t \sim \mathcal{U}(0, 1)} \left[
  \lambda(t) \, \mathbb{E}_{\mathbf{x}_0 \sim q, \,
    \mathbf{x}_t \sim p_{t|0}(\mathbf{x}_t | \mathbf{x}_0)} w(E) \left\|
  s_\theta(\mathbf{x}_t, t) - \nabla_{\mathbf{x}_t} \log
  p_{t|0}(\mathbf{x}_t | \mathbf{x}_0) \right\|_2^2 \right],
\end{equation}
where $\mathbf{x}_0$ now represents buffer samples. The loss emphasizes low-energy structures
without requiring force labels or sampling from the true Boltzmann
distribution. 

\paragraph{Annealing strategy.}
To balance exploration and exploitation, the temperature $T$ is
annealed from a high initial value to a low final value over
training. This encourages broad exploration early on and convergence
toward deep minima in later stages, analogous to simulated annealing.

\paragraph{Force-field guidance.}
Atomic forces are typically available alongside energies at negligible
additional cost. Although they are not used directly in the
Boltzmann-weighted loss, we exploit them through a force-prediction
head attached to the shared representation backbone of the score
network. The predicted forces $F_\theta(\mathbf{x})$
are incorporated into a predictor–corrector sampling scheme: each
reverse diffusion (predictor) step is followed by a force-based
correction step, 
\begin{equation}
  \Delta \mathbf{x} = \alpha (1-t)^\zeta F_\theta(\mathbf{x}),
\end{equation}
where $t$ denotes the diffusion time, $\zeta$ is a scalar hyperparameter, and 
$\alpha$ is an adaptive step size. As $t \rightarrow 0$, the corrective
term plays a progressively larger guiding role, steering samples
toward low-energy configurations, while early in sampling the
stochastic diffusion dynamics promote exploration. The force head is
trained jointly on all evaluated configurations during GO-Diff
training. This approach, similar in spirit to Ref.~\cite{ronne2024},
provides physically grounded force-field guidance (FFG) that
accelerates convergence toward equilibrium configuration.




\paragraph{Diffusion process.}
We follow the methodology of Ref.~\cite{ronne2024,ronne2025} for the
atomistic diffusion process.

\section{Results}
\label{sec:results}
We evaluate GO-Diff on two atomistic optimization tasks using the
MACE-MP0 universal potential\cite{batatia2023}, following the established
benchmark systems from Ref.~\cite{garridotorres2019}. The first task involves
optimizing the placement of a Pt addatom on a stepped Pt
surface. Although the system is three-dimensional, the energy
landscape can be effectively visualized by projecting optimized
addatom positions along the surface normal, yielding a two-dimensional
representation. The second task targets the discovery of a stable
Pt-heptamer cluster on a large $6 \times 6$ Pt(111) surface. This more
complex system serves both as a benchmark against classical RSS  and
as a testbed for amortized optimization via 
transfer of the pretrained score model from the first task. 

\paragraph{Pt-addatom on stepped Pt surface.}

\begin{figure}[h]
\centering
\includegraphics[]{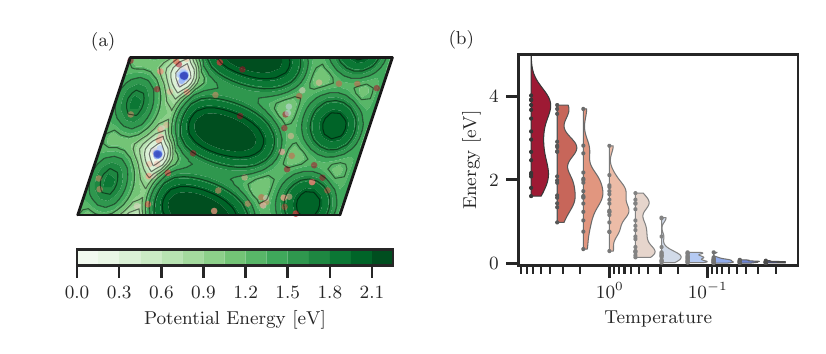}
\caption{\textbf{Addatom optimization on a stepped Pt surface.}
(a) Projected 2D energy surface for different addatom
  placements. Colored dots indicate buffer samples, with color
  denoting annealing temperature (red = high, blue = low). 
(b) Evolution of buffer structure energies during annealing. Sampling
  gradually concentrates in low-energy basins.}
  \label{fig:stepped}
\end{figure}

Figure~\ref{fig:stepped}(a) shows the projected 2D potential energy
surface for different Pt addatom placements. The colored points
represent GO-Diff buffer samples at various temperatures during
annealing. Initially, samples from the untrained model are broadly
scattered. As training progresses and temperature decreases, samples
increasingly concentrate around low-energy basins near the global
minimum. 

Figure~\ref{fig:stepped}(b) illustrates the corresponding energy
evolution of buffer structures. Lower temperatures bias sampling
toward deeper minima, validating the effectiveness of the
Boltzmann-weighted loss and annealing schedule.

\paragraph{Pt-heptamer on Pt(111).}
We next evaluate GO-Diff (with and without FFG) on the more challenging task of discovering
the Pt-heptamer cluster on a $6 \times 6$ Pt(111) surface. We compare
against RSS by measuring both success and the average number of
evaluations required to find the target structure. In addition, we
test amortized optimization: transferring a pretrained GO-Diff model
from the addatom task to initialize the score model in this new
system.

\begin{figure}[h]
\centering
\includegraphics[]{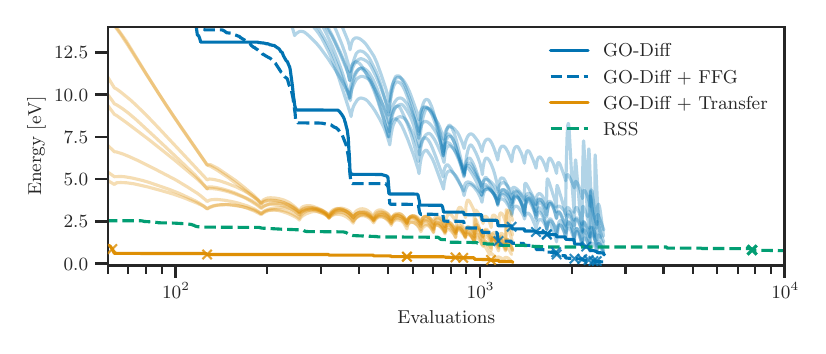}
\caption{\textbf{Benchmarking GO-Diff on Pt-heptamer discovery.}
Mean best energy over eight
independent experiments with faint lines showing the moving average evaluate energy of
each run (Only plotted for GO-Diff and GO-Diff + Transfer). Crosses indicate discovery of the
Pt-heptamer. GO-Diff rapidly improves, while RSS stagnates. Transfer
learning further accelerates convergence.}
\label{fig:heptamer}
\end{figure}

Figure~\ref{fig:heptamer} reports the mean best energy achieved over time
for each method. RSS shows limited progress, often failing to improve
beyond its initial state. In contrast, GO-Diff consistently discovers
the Pt-heptamer within 2,560 energy evaluations. We observe an
improvement using force-field guidance both in terms of evaluated
energies and success. With model transfer,
amortized GO-Diff (without FFG) achieves even faster convergence and better 
performance, finding the target in seven out of eight runs
using $\sim$600 energy evaluation on average. See
Table~\ref{tab:success-metrics} for details.

This acceleration is expected: the transferred model already captures
bonding preferences, such as the stability of hollow sites below step
edges. As a result, the optimization task reduces to adjusting
interatomic geometry, rather than learning bonding from scratch. These
results highlight GO-Diff’s ability to reuse generative knowledge
across systems, enabling amortized optimization in more complex
scenarios.

\begin{table}[h]
\centering
\caption{Comparison of methods by total single-point evaluations, success, and
  mean success iteration. A successful optimization run discovers the Pt-heptamer
  structure.} 
\label{tab:success-metrics}
\begin{tabular}{lccc}
\toprule
\textbf{Method} & \textbf{\# Evaluations} & \textbf{Success} & \textbf{Mean Success Iteration} \\
\midrule
RSS             & 10,000     & 1/8     & 7,816 \\
GO-Diff         & 2,560      & 5/8     & 1,667 \\
GO-Diff + FFG   & 2,560      & \textbf{8/8} & 1,994 \\
GO-Diff + Transfer & \textbf{1,280}   & 7/8     & \textbf{591} \\
\bottomrule
\end{tabular}
\end{table}

Table~\ref{tab:success-metrics} summarizes these findings. GO-Diff
achieves better success across all experiments using fewer evaluations than
RSS, and amortized optimization via model transfer further reduces the
computational budget by more than $2\times$ on average. 

The stochasticity of the diffusion process allows GO-Diff to robustly
escape local minima and explore high-quality regions of the potential
energy surface. Unlike greedy optimizers, it maintains diversity
throughout sampling, guided by the annealed Boltzmann-weighted
objective—enabling both efficiency and reliability even in complex
optimization tasks. 

\section{Discussion}
\label{sec:discussion}

\paragraph{Extending to multi-objective or compositional design.}
A unique strength of GO-Diff is that it can be extended to optimize
over additional properties, such as the atomic composition, electronic
or catalytic properties. These can be incorporated directly into the
diffusion model or via conditioning guidance, enabling optimization
beyond geometry. This is particularly
attractive in materials design, where properties are tightly coupled.


\section{Limitations}
\paragraph{Samples, temperature profile and training steps.}
The number of samples per iteration, the annealing schedule and the
number of training steps per iteration are
critical hyperparameters in GO-Diff. Too few samples reduce buffer
diversity and slow convergence, while overly large batches raise
computational cost with diminishing returns. Likewise, overly
aggressive temperature decay or overfitting each iteration can cause premature collapse around
suboptimal minima, whereas slower schedules and fewer training steps enable broader exploration
and better generalization. In our experiments, we adopt an exponential
decay schedule to balance exploration and exploitation. Future
extensions may include dynamic temperature adjustment, adaptive
sampling, and tuning the number of training steps per iteration—all of
which could improve efficiency and robustness without manual tuning.

\paragraph{Scalability.} The scalability of atomistic diffusion models remains
limited. In practice, current score-based models have mostly been
demonstrated on systems with fewer than 20
atoms.\cite{xie2022,zeni2025} Although recent work such
as~\cite{ronne2024,cornet2024a,joshi2025,cornet2025} has extended diffusion-based approaches to larger
systems, further research is needed to make GO-Diff scalable to
very large and realistic sized systems.

\section{Conclusion}
We introduced GO-Diff, a generative diffusion-based framework for
global structure optimization that avoids relaxation steps and
pretraining. Trained directly on the energy landscape via a
Boltzmann-weighted loss, GO-Diff efficiently samples low-energy
configurations. It outperforms random structure search in both
efficiency and success, and enables amortized optimization
through transfer across related systems. Our results establish
diffusion models as scalable optimization engines for atomistic
modeling, with promising extensions to compositional design, surrogate
acceleration, and multi-objective optimization.

\section*{Acknowledgments}
We are thankful for funding by the Pioneer Center for Accelerating P2X Materials Discovery
(CAPeX), DNRF grant number P3. A.B. thank the Det Frie Forskningsråd
under Project “Data-driven quest for TWh scalable Na-ion battery
(TeraBatt)” (Ref. Number 2035-00232B) and “Autonomous agents of
Discovery for earth-Abundant Na-ion battery cathodes (ADANA)”
(Ref. Number 3164-00297B). The authors also thank the Novo Nordisk
Foundation Data Science Research Infrastructure 2022 Grant: A
high-performance computing infrastructure for data-driven research on
sustainable energy materials, Grant no. NNF22OC0078009. 

\bibliography{bib}

\newpage
\appendix
\section{Appendix}

\subsection{Related work}

\paragraph{Diffusion Models for Black‑Box Optimization.}   
Krishnamoorthy et al.\cite{krishnamoorthy2023} introduce Denoising Diffusion Optimization
Models (DDOM), an inverse approach for black‑box optimization
that learns a conditional diffusion-based generative model mapping
target function values to input configurations. It employs dataset
reweighing and classifier‑free guidance alongside a two-stage
training approach. 

\paragraph{Diffusion Model for Data‑Driven Black‑Box Optimization.}   
Li et al.\cite{li2024} propose a reward‑directed conditional diffusion model
trained on mixed unlabeled and labeled data. By conditioning on high
predicted reward, they cast design optimization as conditional
sampling. 

\paragraph{Iterated Denoising Energy Matching for Sampling from Boltzmann Densities.}
Akhound‑Sadegh et al.\cite{akhound-sadegh2024} introduce iDEM, a novel
diffusion‑based Boltzmann sampler trained using an energy-matching loss
using Monte-Carlo samples to estimate the score. iDEM alternates between drawing
samples from its current model and updating via an energy‑matching
loss. iDEM does not require prior data, and purely relies on energy
and forces of the potential. 

\paragraph{BNEM: A Boltzmann Sampler Based on Bootstrapped Noised Energy Matching.}   
This recent follow‑up\cite{ouyang2025a} extends iDEM by leveraging bootstrapped energy
estimates to improve sampling robustness and improve performance.

\paragraph{Adjoint Sampling: Highly Scalable Diffusion Samplers via Adjoint Matching.}
Havens et al.\cite{havens2025} propose an adjoint‑matching based sampling scheme to
train diffusion-based Boltzmann samplers. The key contribution alongside
the reciprocal adjoint matching is the possibility to get many
gradient updates with few potential energy evaluations and model
samples. They show state-of-the-art performance on synthetic energy
functions and difficult conformational sampling. 

Unlike prior Boltzmann samplers such as iDEM\cite{akhound-sadegh2024} and BNEM\cite{ouyang2025a}, which
estimate training targets from Monte Carlo generated samples, or
Adjoint Sampling\cite{havens2025}, which relies on adjoint matching to improve
update efficiency, GO-Diff uses a direct Boltzmann-weighted
score-matching loss requiring only energy evaluations — avoiding force
labels or MC estimation. Combined with an
annealed temperature schedule and replay-buffer self-sampling, this
yields a simpler and very sample-efficient training loop. Furthermore,
while previous samplers have not demonstrated transfer across systems,
we show that pretrained GO-Diff models can be reused for amortized
global optimization achieving faster optimization.

\subsection{Score-Based Diffusion Models}
Score-based diffusion models\cite{song2021} are a class of generative models that learn to
reverse a diffusion process that progressively adds noise to
data. These models are grounded in stochastic differential equations
(SDEs) and denoising score matching. 

The forward process is defined as a stochastic differential
equation (e.g., a variance-preserving or variance-exploding SDE) that
transforms a data sample $\mathbf{x}_0 \sim p_{\text{data}}$ into a
noisy sample $\mathbf{x}_t$ over time $t \in [0, 1]$. The generative
goal is to model the reverse-time dynamics using a parameterized score
function $s_\theta(\mathbf{x}_t, t) \approx \nabla_{\mathbf{x}_t} \log p_t(\mathbf{x}_t)$. 

The model is trained using score matching, which minimizes the
expected squared difference between added noise to samples and the
prediction hereof. The learning objective is
expressed as:  

\begin{equation}
\mathcal{L}_\theta = \mathbb{E}_{t \sim \mathcal{U}(0, 1)} \left[
  \lambda(t) \, \mathbb{E}_{\mathbf{x}_0 \sim p_{\text{data}}, \,
    \mathbf{x}_t \sim p_{t|0}(\mathbf{x}_t | \mathbf{x}_0)} \left\|
  s_\theta(\mathbf{x}_t, t) - \nabla_{\mathbf{x}_t} \log
  p_{t|0}(\mathbf{x}_t | \mathbf{x}_0) \right\|_2^2 \right], 
\end{equation}
where $\lambda(t)$ is a weighting function depending on the noise
schedule, and $p_{t|0}$ denotes the marginal of the forward SDE
conditioned on the initial data point. 

Once trained, new samples are generated by solving the reverse-time
SDE using numerical solvers such as the Euler-Maruyama samplers.

For a complete description of applying diffusion models to the
materials domain see Ref.~\cite{ronne2025} and specifically for the
application to surface-supported systems see Ref.~\cite{ronne2024}.

\subsection{Buffer update}
We maintain the buffer by applying weighted reservoir
sampling~\cite{efraimidis2006} over all evaluated structures, using
Boltzmann weights (at the current temperature) as sampling
probabilities. This yields a dynamic buffer that gradually shifts
focus toward lower-energy configurations as training progresses.

\subsection{Training algorithm}
Below we present pseudocode for the full GO-Diff optimization loop:

\begin{algorithm}[H]
\caption{GO-Diff Optimization Procedure}
\begin{algorithmic}[1]
\State \textbf{Initialize:} buffer $\mathcal{B} \leftarrow \emptyset$, score model $s_\theta$, temperature schedule $\{T_k\}_{k=1}^K$
\For{iteration $k = 1, \dots, K$}
    \State Set temperature $T \leftarrow T_k$
    \State Sample $N$ structures $\{X^{(j)}\}_{j=1}^N$ from $p_\theta(\mathbf{x}_0)$ via reverse SDE
    \For{each structure $X^{(j)}$}
        \State Evaluate energy $E^{(j)} \leftarrow U(X^{(j)})$
    \EndFor
    \State Update buffer $\mathcal{B}$ with $\{(X^{(j)}, E^{(j)})\}$
    \State Train $s_\theta$ using $\mathcal{L}^{\text{Boltzmann}}_\theta$ on $\mathcal{B}$
\EndFor
\end{algorithmic}
\end{algorithm}

\subsection{Hyperparameters}
We provide common and specific hyperparameters for all experiments.

\begin{table}[h!]
\centering
\caption{Common hyperparameters used in GO-Diff experiments.}
\label{tab:common_hyperparams}
\begin{tabular}{ll}
\toprule
\textbf{Parameter} & \textbf{Value} \\
\midrule
Diffusion sampling steps & 500 \\
Noise schedule & Linear (VE-SDE) \\
Score model architecture & PaiNN GNN (4 blocks);
$6$Å cutoff\\
Final temperature $T_K$ & 0.02 \\
Learning rate & $10^{-4}$ \\
Optimizer & AdamW \\
\bottomrule
\end{tabular}
\end{table}

\begin{table}[h!]
\centering
\caption{Hyperparameters used in \textbf{Pt-addatom on stepped Pt surface} experiment.}
\label{tab:common_hyperparams}
\begin{tabular}{ll}
\toprule
\textbf{Parameter} & \textbf{Value} \\
\midrule
Buffer size $|\mathcal{B}|$ & 16 \\
Samples per iteration $N$ & 32 \\
Initial temperature $T_1$ & 5.0 \\
Annealing schedule & Exponential decay over 10 iterations \\
Batch size & 8 \\
Training epochs per iteration & 1000 \\
\bottomrule
\end{tabular}
\end{table}

\begin{table}[h!]
\centering
\caption{Hyperparameters used in \textbf{Pt-heptamer on Pt(111)} experiment.}
\label{tab:common_hyperparams}
\begin{tabular}{ll}
\toprule
\textbf{Parameter} & \textbf{Value} \\
\midrule
Buffer size $|\mathcal{B}|$ & 64 \\
Samples per iteration $N$ & 128 \\
Initial temperature $T_1$ & 5.0 \\
Annealing schedule & Exponential decay over 20 iterations \\
Batch size & 16 \\
Training epochs per iteration & 2000 \\
\bottomrule
\end{tabular}
\end{table}

The \textbf{Pt-heptamer on Pt(111) + FFG} experiment uses same settings as
without FFG and with $\zeta=3$. The step size $\alpha$ is found using
L-BFGS with a scaling factor of $0.2$. The number of reverse diffusion
steps is kept at 500, but this effectively doubles the number of
model predictions during sampling due to the direct force predictions
for the corrector step. 

\begin{table}[h!]
\centering
\caption{Hyperparameters used in \textbf{Pt-heptamer on Pt(111) + transfer} experiment.}
\label{tab:common_hyperparams}
\begin{tabular}{ll}
\toprule
\textbf{Parameter} & \textbf{Value} \\
\midrule
Buffer size $|\mathcal{B}|$ & 64 \\
Samples per iteration $N$ & 64 \\
Initial temperature $T_1$ & 3.0 \\
Annealing schedule & Exponential decay over 20 iterations \\
Batch size & 16 \\
Training epochs per iteration & 4000 \\
\bottomrule
\end{tabular}
\end{table}

\subsection{RSS}
RSS is performed using the software package
AGOX\cite{christiansen2022} following their documentation.

\subsection{Compute resources}
All experiments where run on single NVIDIA SM3090 GPU with 24GB of
memory.

\subsection{Code availability}
All details to reproduce the experiments are provided at
\url{https://github.com/nronne/go-diff} including specific
implementation details and hyperparameters. The
diffusion model is implemented in the AGeDi software package\cite{ronne2025}.

\end{document}